\documentclass[lettersize,journal]{IEEEtran}
\usepackage{amsmath, amssymb, amsfonts, amsthm, thmtools, algorithmic, algorithm, empheq, lipsum, nicefrac, array}
\usepackage[caption=false,font=normalsize,labelfont=sf,textfont=sf]{subfig}
\usepackage{textcomp, enumitem}
\usepackage{verbatim}
\usepackage{stfloats, graphicx, xcolor}
\usepackage{cite, url}
\usepackage[english]{babel}
\usepackage{float}
\usepackage[caption=false,font=normalsize,labelfont=sf,textfont=sf]{subfig}
\usepackage{caption}
\usepackage{ifpdf}
\usepackage[utf8]{inputenc}
\usepackage{changepage, threeparttable, tabularx}

\usepackage{xr}

\newcommand{\blue}{}

\begin{document}

\title{Performance Bounds for Passive Sensing in Asynchronous ISAC Systems -- Appendices}

\author{Jingbo~Zhao,~
	Zhaoming~Lu,~
	J. Andrew Zhang,~\IEEEmembership{Senior Member,~IEEE,} 
	Weicai Li,~ 
	Yifeng Xiong,~\\
	Zijun Han,~
	Xiangming Wen,~\IEEEmembership{Senior Member,~IEEE}
	and~Tao Gu,~\IEEEmembership{Fellow,~IEEE}
        \IEEEcompsocitemizethanks{
        	\IEEEcompsocthanksitem This document is a supplement to the paper ``Performance Bounds for Passive Sensing in Asynchronous ISAC Systems."
        	\IEEEcompsocthanksitem This work was supported in part by Beijing Natural Science Foundation under Grant JQ21035, in part by National Key R\&D Program of China under Grant 2022YFB2601803. \emph{(Corresponding author: Zhaoming Lu)}
        	\IEEEcompsocthanksitem Jingbo Zhao, Zhaoming Lu, Weicai Li, Yifeng Xiong, Zijun Han and Xiangming Wen are with the School of Information and Communication Engineering, Beijing University of Posts and Telecommunications, Beijing, China. (e-mail: \{zhjb, lzy0372, liweicai, yifengxiong, zijunhan, xiangmw\}@bupt.edu.cn).
        	\IEEEcompsocthanksitem J. Andrew Zhang is with the School of Electrical and Data Engineering, University of Technology Sydney. (e-mail: andrew.zhang@uts.edu.au)
        	\IEEEcompsocthanksitem Tao Gu is with the School of Computing, Macquarie University. (e-mail: tao.gu@mq.edu.au)
        }
    }

\maketitle

\begin{abstract}
	This document contains the appendices for our paper titled ``Performance Bounds for Passive Sensing in Asynchronous ISAC Systems." The appendices include rigorous derivations of key formulas, detailed proofs of the theorems and propositions introduced in the paper, and details of the algorithm tested in the numerical simulation for validation. These appendices aim to support and elaborate on the findings and methodologies presented in the main text. All external references to equations, theorems, and so forth, are directed towards the corresponding elements within the main paper. 
\end{abstract}

\begin{appendices}
\renewcommand{\theequation}{\thesection.\arabic{equation}}

\section{\blue{Proof of CSI as a Sufficient Statistic for Path Parameters}}
\label{Appendix_proof_of_sufficient_statistic}
\setcounter{equation}{0}
\blue{Consider a MIMO-OFDM ISAC system with $M_{\mathrm T}$ transmitting antennas, $M_{\mathrm R}$ receiving antennas, and $K$ subcarriers. Focusing on the $k$-th subcarrier, the system transmit a reference signal (training sequence) $X_k\in \mathbb{C}^{M_{\mathrm T}\times N}, (N\ge M_{\mathrm T})$ to acquire the corresponding CSI estimate $\hat{H}_k\in\mathbb{C}^{R\times T}$ using the least square approach, where $N$ represents the number of transmitted symbols of the reference signal on each subcarrier. 
Assume a semi-unitary reference signal matrix, $X_k X_k^{\dagger} = \boldsymbol{I}$, to achieve the optimum channel estimation performance \cite{Appendix_channel_estimation_identity}.
The total estimated CSI tensor is a stack of $\hat{H}_k$ across all $K$ subcarriers, denoted as $\hat{\boldsymbol{H}}\in\mathbb{C}^{M_{\mathrm R}\times M_{\mathrm T}\times K}$. Similarly, the actual channel state tensor is a stack of the actual channel state matrices $H_k(\Theta)$ across all $K$ subcarriers, denoted as $\boldsymbol{H}(\Theta)\in\mathbb{C}^{M_{\mathrm R}\times M_{\mathrm T}\times K}$, parameterized by the path parameter set $(\Theta)$. The system uses the estimated CSI $\hat{\boldsymbol{H}}$ as one of the snapshots to conduct sensing application, via explicitly estimating the path parameter set $\Theta$ or extracting feature signals using machine learning techniques.}

\blue{The received signal on the $k$-the subcarrier is given by $Y_k \in \mathbb{C}^{M_{\mathrm R}\times N}$,
\begin{equation}
	Y_k = H_k(\Theta) \cdot X_k + \boldsymbol{n},
\end{equation}
where $\boldsymbol{n}$ is the i.i.d. complex Gaussian noise on the raw received signal. We also express the total received signal as the stack of $Y_k$ across all subcarriers, denoted as $\boldsymbol{Y}\in\mathbb{C}^{M_{\mathrm R}\times N\times K}$. The corresponding CSI estimate on the $k$-th subcarrier by the LS approach is given by
\begin{equation}
	\hat{H}_k(Y_k) = Y_k \cdot X_k^{\dagger},
\end{equation}
and we express the total CSI estimate tensor by stacking $\hat{H}_k(Y_k)$ across all $K$ subcarriers, denoted as $\hat{\boldsymbol{H}} (\boldsymbol{Y})$. }

\begin{figure*}[b]
	\hrulefill
	\blue{
	\begin{equation}
		\label{Equ_pdf}
		\begin{aligned}
			p_{\boldsymbol{Y}}(\boldsymbol{Y};\Theta) 
			=& \prod_{k=1}^K \frac{1}{\sqrt{(2\pi)^{M_{\mathrm R}N}M_{\mathrm R}N}}\exp\!\!\bigg(\!\!\!-\!\frac{1}{2}\mathrm{tr}\!\Big[ \big(Y_k - H_k(\Theta)\cdot X_k\big) \big(Y_k - H_k(\Theta)\cdot X_k\big)^{\dagger} \Big] \!\bigg)\\
			=& \underbrace{\frac{1}{\sqrt{(2\pi)^{M_{\mathrm R}N\!K}M_{\mathrm R}^{K}N^{K}}} \exp\!\!\bigg(\!\!\!-\!\frac{1}{2}\sum_{k=1}^{K} \mathrm{tr}\!\Big[Y_k Y_k^{\dagger}\Big]\!\!\bigg)}_{h(\boldsymbol{Y})} 
			\cdot \underbrace{\exp\!\!\Bigg(\!\!\!-\!\frac{1}{2}\sum_{k=1}^{K} \mathrm{tr}\!\Big[H_k(\Theta) H_k^{\dagger}(\Theta)\Big] + \sum_{k=1}^{K} \Re\!\bigg\{\!\mathrm{tr}\!\Big[\hat{H}_k (Y_k) H_k^{\dagger}(\Theta)\Big]\!\!\bigg\}\!\!\Bigg)}_{g(\hat{\boldsymbol{H}} (\boldsymbol{Y}), \Theta)} \!\!.
		\end{aligned}
	\end{equation}}
\end{figure*}

\blue{Upon the assumption that the noise is independent across subcarriers, The probability density function of the total received signal $\boldsymbol{Y}$ is given by \eqref{Equ_pdf} at the bottom of next page.
According to the Neyman-Fisher factorization theorem \cite{Appendix_fundamentals}, one may now see that the CSI, $\hat{\boldsymbol{H}} (\boldsymbol{Y})$, is a sufficient statistic for $\Theta$. }

\section{Expression of the joint FIM}
\setcounter{equation}{0}
\label{appendix_joint_FIM}
From (10) and (11), the joint FIM in the single-carrier model, $\boldsymbol{\mathrm{J}}^{\text{S}, \mathrm{Full}}$, is structured as 
\begin{equation}
\label{equ_J_full}
\begin{aligned}
	&\!\!\!\!\boldsymbol{\mathrm{J}}^{\text{S}, \mathrm{Full}}\!\!=\!\!
	\begin{bmatrix}
			\!\!\!\mathrm{J}^{\text{S}}_{\theta_{\mathrm{d}}\!,\theta_{\mathrm{d}}} &\!\!\!\!\!\!\!
			\boldsymbol{\mathrm{J}}^{\text{S}}_{\theta_{\mathrm{d}}\!,\boldsymbol{h}_{\mathrm{s\!,r}}^{\text{S}}} &\!\!\!\!\!\!\!
			\boldsymbol{\mathrm{J}}^{\text{S}}_{\theta_{\mathrm{d}}\!,\boldsymbol{h}_{\mathrm{s\!,i}}^{\text{S}}} &\!\!\!\!\!\!\!
			\boldsymbol{\mathrm{J}}^{\text{S}}_{\theta_{\mathrm{d}}\!,\boldsymbol{d}_\mathrm{r}}&\!\!\!\!\!\!\!
			\boldsymbol{\mathrm{J}}^{\text{S}}_{\theta_{\mathrm{d}}\!,\boldsymbol{d}_\mathrm{i}}&\!\!\!\!\!\!\!
			\boldsymbol{\mathrm{J}}^{\text{S}}_{\theta_{\mathrm{d}}\!,\boldsymbol{\varphi}_{\mathrm o}}\!\!
			\\
			\!{\boldsymbol{\mathrm{J}}^{\text{S}}_{\theta_{\mathrm{d}}\!,\boldsymbol{h}^{\text{S}}_{\mathrm{s\!,r}}}}^{\!\!\!\!\intercal} &\!\!\!\!\!
			\boldsymbol{\mathrm{J}}^{\text{S}}_{\boldsymbol{h}_{\mathrm{s\!,r}}^{\text{S}}\!,\boldsymbol{h}_{\mathrm{s\!,r}}^{\text{S}}} &\!\!\!\!\!
			\boldsymbol{\mathrm{J}}^{\text{S}}_{\boldsymbol{h}_{\mathrm{s\!,r}}^{\text{S}}\!,\boldsymbol{h}_{\mathrm{s\!,i}}^{\text{S}}} &\!\!\!\!\!
			\boldsymbol{\mathrm{J}}^{\text{S}}_{\boldsymbol{h}_{\mathrm{s\!,r}}^{\text{S}}\!,\boldsymbol{d}_\mathrm{r}} &\!\!\!\!\!
			\boldsymbol{\mathrm{J}}^{\text{S}}_{\boldsymbol{h}_{\mathrm{s\!,r}}^{\text{S}}\!,\boldsymbol{d}_\mathrm{i}} &\!\!\!\!\!
			\boldsymbol{\mathrm{J}}^{\text{S}}_{\boldsymbol{h}_{\mathrm{s\!,r}}^{\text{S}}\!,\boldsymbol{\varphi}_{\mathrm o}}\!\!
			\\
			\!{\boldsymbol{\mathrm{J}}^{\text{S}}_{\theta_{\mathrm{d}}\!,\boldsymbol{h}^{\text{S}}_{\mathrm{s\!,i}}}}^{\!\!\!\!\intercal} &\!\!\!\!\!
			{\boldsymbol{\mathrm{J}}^{\text{S}}_{\boldsymbol{h}_{\mathrm{s\!,r}}^{\text{S}}\!,\boldsymbol{h}_{\mathrm{s\!,i}}^{\text{S}}}}^{\!\!\!\!\intercal} &\!\!\!\!\!
			\boldsymbol{\mathrm{J}}^{\text{S}}_{\boldsymbol{h}_{\mathrm{s\!,i}}^{\text{S}}\!,\boldsymbol{h}_{\mathrm{s\!,i}}^{\text{S}}} &\!\!\!\!\!
			\boldsymbol{\mathrm{J}}^{\text{S}}_{\boldsymbol{h}_{\mathrm{s\!,i}}^{\text{S}}\!,\boldsymbol{d}_\mathrm{r}} &\!\!\!\!\!
			\boldsymbol{\mathrm{J}}^{\text{S}}_{\boldsymbol{h}_{\mathrm{s\!,i}}^{\text{S}}\!,\boldsymbol{d}_\mathrm{i}} &\!\!\!\!\!
			\boldsymbol{\mathrm{J}}^{\text{S}}_{\boldsymbol{h}_{\mathrm{s\!,i}}^{\text{S}}\!,\boldsymbol{\varphi}_{\mathrm o}}\!\!
			\\
			\!\!\!{\boldsymbol{\mathrm{J}}^{\text{S}}_{\theta_{\mathrm{d}}\!,\boldsymbol{d}_\mathrm{r}}}^{\!\!\!\!\intercal}&\!\!\!\!\!\!\!
			{\boldsymbol{\mathrm{J}}^{\text{S}}_{\theta_{\mathrm{d}}\!,\boldsymbol{h}_{\mathrm{s,r}}^{\text{S}}}}^{\!\!\!\!\intercal} &\!\!\!\!\!\!
			{\boldsymbol{\mathrm{J}}^{\text{S}}_{\theta_{\mathrm{d}}\!,\boldsymbol{h}_{\mathrm{s,i}}^{\text{S}}}}^{\!\!\!\!\intercal} &\!\!\!\!\!\!\!\!
			\boldsymbol{\mathrm{J}}^{\text{S}}_{\boldsymbol{d}_\mathrm{r}\!, \boldsymbol{d}_\mathrm{r}}&\!\!\!\!\!\!\!
			\boldsymbol{\mathrm{J}}^{\text{S}}_{\boldsymbol{d}_\mathrm{r}\!, \boldsymbol{d}_\mathrm{i}}&\!\!\!\!\!\!\!
			\boldsymbol{\mathrm{J}}^{\text{S}}_{\boldsymbol{d}_\mathrm{r}\!,\boldsymbol{\varphi}_{\mathrm o}}\!\!
			\\
			\!\!\!\!{\boldsymbol{\mathrm{J}}^{\text{S}}_{\theta_{\mathrm{d}}\!,\boldsymbol{d}_\mathrm{i}}}^{\!\!\!\!\intercal}&\!\!\!\!\!\!\!\!
			{\boldsymbol{\mathrm{J}}^{\text{S}}_{\theta_{\mathrm{d}}\!,\boldsymbol{h}_{\mathrm{s\!,r}}^{\text{S}}}}^{\!\!\!\!\intercal} &\!\!\!\!\!\!\!\!
			{\boldsymbol{\mathrm{J}}^{\text{S}}_{\theta_{\mathrm{d}}\!,\boldsymbol{h}_{\mathrm{s\!,i}}^{\text{S}}}}^{\!\!\!\!\intercal} &\!\!\!\!\!\!\!\!
			{\boldsymbol{\mathrm{J}}^{\text{S}}_{\boldsymbol{d}_\mathrm{r}, \boldsymbol{d}_\mathrm{i}}}^{\!\!\!\!\intercal}&\!\!\!\!\!\!\!\!
			\boldsymbol{\mathrm{J}}^{\text{S}}_{\boldsymbol{d}_\mathrm{i}\!, \boldsymbol{d}_\mathrm{i}}&\!\!\!\!\!\!\!\!
			\boldsymbol{\mathrm{J}}^{\text{S}}_{\boldsymbol{d}_\mathrm{i}\!,\boldsymbol{\varphi}_{\mathrm o}}\!\!
			\\
			\!\!\!\!{\boldsymbol{\mathrm{J}}^{\text{S}}_{\theta_{\mathrm{d}}\!,\boldsymbol{\varphi}_{\mathrm o}}}^{\!\!\!\!\intercal}&\!\!\!\!\!\!\!\!
			{\boldsymbol{\mathrm{J}}^{\text{S}}_{\theta_{\mathrm{d}}\!,\boldsymbol{h}_{\mathrm{s\!,r}}^{\text{S}}}}^{\!\!\!\!\intercal} &\!\!\!\!\!\!\!\!
			{\boldsymbol{\mathrm{J}}^{\text{S}}_{\theta_{\mathrm{d}}\!,\boldsymbol{h}_{\mathrm{s\!,i}}^{\text{S}}}}^{\!\!\!\!\intercal} &\!\!\!\!\!\!\!\!
			{\boldsymbol{\mathrm{J}}^{\text{S}}_{\boldsymbol{d}_\mathrm{r}\!,\boldsymbol{\varphi}_{\mathrm o}}}^{\!\!\!\!\intercal}&\!\!\!\!\!\!\!\!
			{\boldsymbol{\mathrm{J}}^{\text{S}}_{\boldsymbol{d}_\mathrm{i}\!,\boldsymbol{\varphi}_{\mathrm o}}}^{\!\!\!\!\intercal}&\!\!\!\!\!\!\!\!
			\boldsymbol{\mathrm{J}}^{\text{S}}_{\boldsymbol{\varphi}_{\mathrm o}\!,\boldsymbol{\varphi}_{\mathrm o}}\!\!
		\end{bmatrix}\!\!,\!\!\!\!\!\!
\end{aligned}
\end{equation}
where the submatrices are given by
{\allowdisplaybreaks
	\begin{align} 
		\label{FIM_entries}
		&\mathrm{J}^{\text{S}}_{\theta_{\mathrm{d}},\theta_{\mathrm{d}}}=\left| \boldsymbol{b}_{\theta}(\theta_{\mathrm{d}})\right|^2 \left| \boldsymbol{d}\right|^2 \nonumber\\
		&\boldsymbol{\mathrm{J}}^{\text{S}}_{\theta_{\mathrm{d}},\boldsymbol{h}^{\text{S}}_{\mathrm{s,r}}}=\Re\left\{ ({\boldsymbol{d}}^{\dagger}\boldsymbol{1}_{T \times 1}) \cdot {\boldsymbol{b}_{\theta}(\theta_{\mathrm{d}})}^{\dagger} \right\} \nonumber\\
		&\boldsymbol{\mathrm{J}}^{\text{S}}_{\theta_{\mathrm{d}},\boldsymbol{h}^{\text{S}}_{\mathrm{s,i}}}=
		-\Im\left\{ ({\boldsymbol{d}}^{\dagger}\boldsymbol{1}_{T \times 1}) \cdot {\boldsymbol{b}_{\theta}(\theta_{\mathrm{d}})}^{\dagger} \right\} \nonumber\\
		&\boldsymbol{\mathrm{J}}^{\text{S}}_{\theta_{\mathrm{d}},\boldsymbol{d}_{\mathrm{r}}}=
		\Re\left\{ \left({\boldsymbol{b}_{\theta}(\theta_{\mathrm{d}})}^{\dagger}\boldsymbol{a}_{\theta}(\theta_{\mathrm{d}})\right) \cdot {\boldsymbol{d}}^{\dagger} \right\} \nonumber\\
		&\boldsymbol{\mathrm{J}}^{\text{S}}_{\theta_{\mathrm{d}},\boldsymbol{d}_{\mathrm{i}}}=
		-\Im\left\{ \left({\boldsymbol{b}_{\theta}(\theta_{\mathrm{d}})}^{\dagger}\boldsymbol{a}_{\theta}(\theta_{\mathrm{d}})\right) \cdot {\boldsymbol{d}}^{\dagger} \right\} \nonumber\\
		&\boldsymbol{\mathrm{J}}^{\text{S}}_{\theta_{\mathrm{d}},\boldsymbol{\varphi}_{\mathrm o}}= \nonumber\\
		&-\Im\left\{ \left({\boldsymbol{b}_{\theta}(\theta_{\mathrm{d}})}^{\dagger}\boldsymbol{h}^{\text{S}}_{\mathrm{s}}\right)\cdot{\boldsymbol{d}}^{\dagger} + \left({\boldsymbol{b}_{\theta}(\theta_{\mathrm{d}})}^{\dagger}\boldsymbol{a}_{\theta}(\theta_{\mathrm{d}})\right) \cdot {\boldsymbol{d}}^{\dagger} \cdot \mathrm{diag}(\boldsymbol{d})\right\}\nonumber\\
		&\boldsymbol{\mathrm{J}}^{\text{S}}_{\boldsymbol{h}^{\text{S}}_{\mathrm{s,r}},\boldsymbol{h}^{\text{S}}_{\mathrm{s,r}}}=
		T\cdot \boldsymbol{\mathrm{I}}_{M\times M}\nonumber\\
		&\boldsymbol{\mathrm{J}}^{\text{S}}_{\boldsymbol{h}^{\text{S}}_{\mathrm{s,r}},\boldsymbol{h}^{\text{S}}_{\mathrm{s,i}}}=
		\boldsymbol{0}\nonumber\\
		&\boldsymbol{\mathrm{J}}^{\text{S}}_{\boldsymbol{h}^{\text{S}}_{\mathrm{s,r}},\boldsymbol{d}_{\mathrm{r}}}=
		\Re\left\{ \boldsymbol{a}_{\theta}(\theta_{\mathrm{d}}) \cdot \boldsymbol{1}_{1\times T} \right\}\nonumber\\
		&\boldsymbol{\mathrm{J}}^{\text{S}}_{\boldsymbol{h}^{\text{S}}_{\mathrm{s,r}},\boldsymbol{d}_{\mathrm{i}}}=
		-\Im\left\{ \boldsymbol{a}_{\theta}(\theta_{\mathrm{d}}) \cdot \boldsymbol{1}_{1\times T} \right\}\nonumber\\
		&\boldsymbol{\mathrm{J}}^{\text{S}}_{\boldsymbol{h}^{\text{S}}_{\mathrm{s,r}},\boldsymbol{\varphi}_{\mathrm o}}=
		-\Im\left\{ \boldsymbol{h}^{\text{S}}_{\mathrm{s}} \cdot \boldsymbol{1}_{1 \times T} + \boldsymbol{a}_{\theta}(\theta_{\mathrm{d}}) \cdot {\boldsymbol{d}}^{\intercal} \right\}\nonumber\\
		&\boldsymbol{\mathrm{J}}^{\text{S}}_{\boldsymbol{h}^{\text{S}}_{\mathrm{s,i}},\boldsymbol{h}^{\text{S}}_{\mathrm{s,i}}}=
		T\cdot \boldsymbol{\mathrm{I}}_{M\times M}\nonumber\\
		&\boldsymbol{\mathrm{J}}^{\text{S}}_{\boldsymbol{h}^{\text{S}}_{\mathrm{s,i}},\boldsymbol{d}_{\mathrm{r}}}=
		\Im\left\{ \boldsymbol{a}_{\theta}(\theta_{\mathrm{d}}) \cdot \boldsymbol{1}_{1\times T} \right\}\nonumber\\
		&\boldsymbol{\mathrm{J}}^{\text{S}}_{\boldsymbol{h}^{\text{S}}_{\mathrm{s,i}},\boldsymbol{d}_{\mathrm{i}}}=
		\Re\left\{ \boldsymbol{a}_{\theta}(\theta_{\mathrm{d}}) \cdot \boldsymbol{1}_{1\times T} \right\}\nonumber\\
		&\boldsymbol{\mathrm{J}}^{\text{S}}_{\boldsymbol{h}^{\text{S}}_{\mathrm{s,i}},\boldsymbol{\varphi}_{\mathrm o}}=
		\Re\left\{ \boldsymbol{h}^{\text{S}}_{\mathrm{s}} \cdot \boldsymbol{1}_{1 \times T} + \boldsymbol{a}_{\theta}(\theta_{\mathrm{d}}) \cdot {\boldsymbol{d}}^{\intercal} \right\}\nonumber\\
		&\boldsymbol{\mathrm{J}}^{\text{S}}_{\boldsymbol{d}_{\mathrm{r}},\boldsymbol{d}_{\mathrm{r}}}=
		M\cdot \boldsymbol{\mathrm{I}}_{T\times T}\nonumber\\
		&\boldsymbol{\mathrm{J}}^{\text{S}}_{\boldsymbol{d}_{\mathrm{r}},\boldsymbol{d}_{\mathrm{i}}}=
		\boldsymbol{0}\nonumber\\
		&\boldsymbol{\mathrm{J}}^{\text{S}}_{\boldsymbol{d}_{\mathrm{r}},\boldsymbol{\varphi}_{\mathrm o}}=
		-\Im\left\{ \mathrm{diag}\left( {\boldsymbol{a}_{\theta}(\theta_{\mathrm{d}})}^{\dagger}\boldsymbol{h}^{\text{S}}_{\mathrm{s}}\cdot\boldsymbol{1}_{T\times 1} + M \cdot\boldsymbol{d} \right) \right\}\nonumber\\
		&\boldsymbol{\mathrm{J}}^{\text{S}}_{\boldsymbol{d}_{\mathrm{i}},\boldsymbol{d}_{\mathrm{i}}}=
		M\cdot \boldsymbol{\mathrm{I}}_{T\times T}\nonumber\\
		&\boldsymbol{\mathrm{J}}^{\text{S}}_{\boldsymbol{d}_{\mathrm{r}},\boldsymbol{\varphi}_{\mathrm o}}=
		-\Re\left\{ \mathrm{diag}\left( {\boldsymbol{a}_{\theta}(\theta_{\mathrm{d}})}^{\dagger}\boldsymbol{h}^{\text{S}}_{\mathrm{s}}\cdot\boldsymbol{1}_{T\times 1} + M \cdot\boldsymbol{d} \right) \right\}\nonumber\\
		&\boldsymbol{\mathrm{J}}^{\text{S}}_{\boldsymbol{\varphi}_{\mathrm o},\boldsymbol{\varphi}_{\mathrm o}}= \Re\Bigg\{ \boldsymbol{\mathrm{I}}_{T\times T} \nonumber\\
		& \odot \!\left(\!\!{\left( \boldsymbol{h}^{\text{S}}_{\mathrm{s}} \!\cdot\! \boldsymbol{1}_{1 \times T} + \boldsymbol{a}_{\theta}(\theta_{\mathrm{d}}) \!\cdot\! {\boldsymbol{d}}^{\intercal} \right)}^{\!\dagger}\!\left( \boldsymbol{h}^{\text{S}}_{\mathrm{s}} \!\cdot\! \boldsymbol{1}_{1 \times T} + \boldsymbol{a}_{\theta}(\theta_{\mathrm{d}}) \!\cdot\! {\boldsymbol{d}}^{\intercal} \right) \!\!\right) \!\!\!\Bigg\} .
	\end{align}}

\section{An example of matrix $\boldsymbol{\mathrm{U}}$ for constrained CRBs}
\setcounter{equation}{0}
	\label{appendix_constrained_CRBs}
	The value of matrix $\boldsymbol{\mathrm{U}}^{\text{S}}$ is not unique. We provide an example of legitimate constructions, 
	\begin{equation}
		\boldsymbol{\mathrm{U}}^{\text{S}} = 
		\begin{bmatrix}
		\boldsymbol{\mathrm{I}}_{(2M+1)\times(2M+1)} &\boldsymbol{0} &\boldsymbol{0} &\boldsymbol{0} \\
		\boldsymbol{0} &\boldsymbol{\mathrm{U}}_{\mathrm{sub}} &\boldsymbol{0} &\boldsymbol{0} \\
		\boldsymbol{0} &\boldsymbol{0} &\boldsymbol{\mathrm{U}}_{\mathrm{sub}} &\boldsymbol{0} \\
		\boldsymbol{0} &\boldsymbol{0} &\boldsymbol{0} &\boldsymbol{\mathrm{U}}_{\mathrm{sub}}
		\end{bmatrix} , 
	\end{equation}
	where $\boldsymbol{\mathrm{U}}_{\mathrm{sub}}$ is given by
	\begin{equation}
		\boldsymbol{\mathrm{U}}_{\mathrm{sub}} = 
		\begin{bmatrix}
			\frac{1}{T+\sqrt{T}}\cdot\boldsymbol{1}_{(T-1)\times (T-1)}- \boldsymbol{\mathrm{I}} \\
			\frac{1}{\sqrt{T}}\cdot\boldsymbol{1}_{1 \times (T-1)} 
		\end{bmatrix}. 
	\end{equation}

\begin{figure*}[b]
	\setcounter{section}{5}
	\setcounter{equation}{0}
	\vspace{-0.3cm}
	\hrulefill
	\vspace{-0.1cm}
	\begin{subequations}
		\label{equ_FIM_reorder_entity}
		\begin{align}
			\label{equ_FIM_reorder_entity:a}
			&\boldsymbol{\mathrm{J}}_{\boldsymbol{\psi}_t^{\text{S}}, \boldsymbol{\psi}_t^{\text{S}}}^{\text{S}} = \frac{1}{\sigma^2} 
			\begin{bmatrix}
				M &0 &-\Im \left\{ {\boldsymbol{a}_{\theta}\!\left(\theta_{\mathrm{d}}\right)^\dagger}\boldsymbol{h}^{\text{S}}_{\mathrm s} + Md_t \right\} \\
				0 & M & \Re \left\{ {{\boldsymbol{a}_{\theta}\!\left(\theta_{\mathrm{d}}\right)^\dagger }\boldsymbol{h}^{\text{S}}_{\mathrm s} + Md_t} \right\} \\
				-\Im \left\{ {\boldsymbol{a}_{\theta}\!\left(\theta_{\mathrm{d}}\right)^\dagger}\boldsymbol{h}^{\text{S}}_{\mathrm s} + Md_t \right\}  & \Re \left\{ {{\boldsymbol{a}_{\theta}\!\left(\theta_{\mathrm{d}}\right)^\dagger }\boldsymbol{h}^{\text{S}}_{\mathrm s} + Md_t} \right\} & 
				\left|\boldsymbol{h}^{\text{S}}_{\mathrm s}+\boldsymbol{a}_{\theta}\!\left(\theta_{\mathrm{d}}\right)d_t\right|^2
			\end{bmatrix}
		\end{align}
		\begin{align}
			\label{equ_FIM_reorder_entity:b}
			\boldsymbol{\mathrm{J}}_{\theta_{\mathrm{d}},\boldsymbol{\psi}_t^{\text{S}}}^{\text{S}}=\frac{1}{\sigma^2}
			\begin{bmatrix} 
				\Re\!\left\{ \boldsymbol{a}_{\theta}\!\left(\theta_{\mathrm{d}}\right)^\dagger \boldsymbol{b}_{\theta}\!\left(\theta_{\mathrm{d}}\right)d_t\right\} &
				\Im\!\left\{ \boldsymbol{a}_{\theta}\!\left(\theta_{\mathrm{d}}\right)^\dagger \boldsymbol{b}_{\theta}\!\left(\theta_{\mathrm{d}}\right)d_t\right\} &
				-\Im\left\{ \boldsymbol{b}_{\theta}\!\left(\theta_{\mathrm{d}}\right)^\dagger \boldsymbol{a}_{\theta}\!\left(\theta_{\mathrm{d}}\right)d^*d+\boldsymbol{b}_{\theta}\!\left(\theta_{\mathrm{d}}\right)^\dagger \boldsymbol{h}_{\mathrm{s}}d_t^*\right\}
			\end{bmatrix}
		\end{align}
	\end{subequations}
	\setcounter{equation}{4}
	\begin{equation} 
		\label{equ_FIM_inverse}
		\begin{aligned}
			&{\boldsymbol{\mathrm{J}}_{\boldsymbol{\psi}_t^{\text{S}},\boldsymbol{\psi}_t^{\text{S}}}^{\text{S}}}^{-1} 
			= \frac{\sigma^2}{M\left(\left|\boldsymbol{a}_{\theta}\!\left(\theta_{\mathrm{d}}\right)\right|^2 \left|\boldsymbol{h}^{\text{S}}_{\mathrm s}\right|^2-\left|\boldsymbol{a}_{\theta}\!\left(\theta_{\mathrm{d}}\right)^\dagger \boldsymbol{h}^{\text{S}}_{\mathrm s}\right|^2\right)} \cdot
			\begin{bmatrix}
				\Im\left\{\chi\right\}^2&\!\!\!-\Re\left\{\chi\right\} \Im\left\{\chi\right\}&\!\!\!M\Im\left\{\chi\right\}\\
				-\Re\left\{\chi\right\} \Im\left\{\chi\right\}&\!\!\!\Re\left\{\chi\right\}^2&\!\!\!-M\Re\left\{\chi\right\}\\
				M\Im\left\{\chi\right\}&\!\!\!-M\Re\left\{\chi\right\}&\!\!\!M^2
			\end{bmatrix}
			+\frac{\sigma^2}{M}\begin{bmatrix} 1&0&0\\0&1&0\\0&0&0 \end{bmatrix},\\
			& \text{where} \, \chi = \boldsymbol{a}_{\theta}\!\left(\theta_{\mathrm{d}}\right)^\dagger \boldsymbol{h}^{\text{S}}_{\mathrm{s}}+Md_t
		\end{aligned}
	\end{equation}
	\begin{equation}
		\label{equ_FIM_theta_equ_result}
		\mathrm{J}_{\theta_{\mathrm{d}}}^\mathrm{S, equ}=\frac{\left|\boldsymbol{d}\right|^2}{\sigma^2 M}
		\left(\!\left|\boldsymbol{a}_{\theta}\!\!\left(\theta_{\mathrm{d}}\right)\right|^2 \!\left|\boldsymbol{b}_{\theta}\!\left(\theta_{\mathrm{d}}\right)\right|^2 \!-\! \left|\boldsymbol{a}_{\theta}\!\!\left(\theta_{\mathrm{d}}\right)^{\!\dagger} \!\boldsymbol{b}_{\theta}\!\left(\theta_{\mathrm{d}}\right)\right|^2 \right) 
		- \frac{ \sum\limits_{t=1}^{K} \Im\bigg\{\!\!\left(
			\boldsymbol{b}_{\theta}\!\left(\theta_{\mathrm{d}}\right)^{\!\dagger}\!\boldsymbol{a}_{\theta}\!\!\left(\theta_{\mathrm{d}}\right)\boldsymbol{a}_{\theta}\!\!\left(\theta_{\mathrm{d}}\right)^{\!\dagger}\!\boldsymbol{h}^{\text{S}}_\mathrm{s} -\boldsymbol{a}_{\theta}\!\left(\theta_{\mathrm{d}}\right)^{\!\dagger}\!\boldsymbol{a}_{\theta}\!\!\left(\theta_{\mathrm{d}}\right)\boldsymbol{b}_{\theta}\!\left(\theta_{\mathrm{d}}\right)^{\!\dagger}\!\boldsymbol{h}^{\text{S}}_\mathrm{s}\right)\! d_t^* \!\bigg\}^2 }
		{\sigma^2 M\left(\left|\boldsymbol{a}_{\theta}\!\left(\theta_{\mathrm{d}}\right)\right|^2 \left|\boldsymbol{h}^{\text{S}}_{\mathrm s}\right|^2-\left|\boldsymbol{a}_{\theta}\!\left(\theta_{\mathrm{d}}\right)^{\!\dagger} \boldsymbol{h}^{\text{S}}_{\mathrm s}\right|^2\right)}
	\end{equation}
	
	\setcounter{section}{3}
	\setcounter{equation}{0}
\end{figure*}

\section{Proof of the Cram\'{e}r-Rao inequality for the HRCRBs}
\setcounter{equation}{0}
	\label{appendix_proof_relaxed_CRB}
	According to the classic Cram\'{e}r-Rao inequality,
	\begin{equation}
		\label{equ_CR_inequality}
		\boldsymbol{C\!ov}_{\boldsymbol{\psi}}  - \boldsymbol{\mathrm{J}}_{\boldsymbol{\psi}}^{-1} \geq \boldsymbol{0},
	\end{equation}
	where $\boldsymbol{C\!ov}_{\boldsymbol{\psi}}$ is the error covariance matrix of any unbiased estimate of $\boldsymbol{\psi}$, 
	\begin{equation}
		\boldsymbol{C\!ov}_{\boldsymbol{\psi}}=
		\begin{bmatrix}
			\boldsymbol{C\!ov}_{\boldsymbol{\psi}_{\mathrm{a}}, \boldsymbol{\psi}_{\mathrm{a}}} &\boldsymbol{C\!ov}_{\boldsymbol{\psi}_{\mathrm{a}},\boldsymbol{\psi}_{\mathrm{b}}} \\
			\boldsymbol{C\!ov}_{\boldsymbol{\psi}_{\mathrm{a}},\boldsymbol{\psi}_{\mathrm{b}}}^{\intercal} &\boldsymbol{C\!ov}_{\boldsymbol{\psi}_{\mathrm{b}}, \boldsymbol{\psi}_{\mathrm{b}}}
		\end{bmatrix}.
	\end{equation}
	Since the leftside of \eqref{equ_CR_inequality} is positive semi-definite, its principal submatrice is also semipositive, i.e.,
	\begin{equation}
		\boldsymbol{C\!ov}_{\boldsymbol{\psi}_{\mathrm{a}}, \boldsymbol{\psi}_{\mathrm{a}}} - \left[\boldsymbol{\mathrm{J}}_{\boldsymbol{\psi}}^{-1}\right]_{1:L_{\mathrm{a}},1:L_{\mathrm{a}}} \geq \boldsymbol{0}
	\end{equation}
	Taking expectation over $\boldsymbol{\psi}_{\mathrm{b}}$ on both sides yields 
	\begin{equation}
		\boldsymbol{C\!ov}_{\boldsymbol{\psi}_{\mathrm{a}}} \geq \left[\mathbb{E}_{\boldsymbol{\psi}_{\mathrm{b}}}\left\{\boldsymbol{\mathrm{J}}_{\boldsymbol{\psi}}^{-1}\right\}\right]_{1:L_{\mathrm{a}},1:L_{\mathrm{a}}} , 
	\end{equation}
	where $\boldsymbol{C\!ov}_{\boldsymbol{\psi}_{\mathrm{a}}}=\mathbb{E}_{\boldsymbol{\psi}_{\mathrm{b}}}\left\{\boldsymbol{C\!ov}_{\boldsymbol{\psi}_{\mathrm{a}}, \boldsymbol{\psi}_{\mathrm{a}}}\right\}$ is the error covariance matrix of any unbiased estimate of $\boldsymbol{\psi}_{\mathrm{a}}$, considering the randomness of $\boldsymbol{\psi}_{\mathrm{b}}$.
	This validates the last inequality of (17). 
	
	The second inequality of (17), 
	\begin{equation}
		\left[\mathbb{E}_{\boldsymbol{\psi}_{\mathrm{b}}}\left\{\boldsymbol{\mathrm{J}}_{\boldsymbol{\psi}}\right\}^{-1}\right]_{1:L_{\mathrm{a}},1:L_{\mathrm{a}}}\leq \left[\mathbb{E}_{\boldsymbol{\psi}_{\mathrm{b}}}\left\{\boldsymbol{\mathrm{J}}_{\boldsymbol{\psi}}^{-1}\right\}\right]_{1:L_{\mathrm{a}},1:L_{\mathrm{a}}}, 
	\end{equation} 
	can be established via Jensen's inequality. 
	
	We now prove the first inequality \mbox{of (17).} 
	Since $\boldsymbol{\mathrm{J}}_{\boldsymbol{\psi}_{\mathrm{b}},\boldsymbol{\psi}_{\mathrm{b}}}$ is positive semi-definite, 
	$\mathbb{E}_{\boldsymbol{\psi}_\mathrm{\!b}}\!\!\!\left\{\!\boldsymbol{\mathrm{J}}_{\boldsymbol{\psi}_\mathrm{\!a}, \boldsymbol{\psi}_\mathrm{\!b}}\!\right\}\! \mathbb{E}_{\boldsymbol{\psi}_\mathrm{\!b}}\!\!\!\left\{\!\boldsymbol{\mathrm{J}}_{\boldsymbol{\psi}_\mathrm{\!b}, \boldsymbol{\psi}_\mathrm{\!b}}\!\right\}^{\!\!-\!1}\!\mathbb{E}_{\boldsymbol{\psi}_\mathrm{\!b}}\!\!\!\left\{\!\boldsymbol{\mathrm{J}}_{\boldsymbol{\psi}_\mathrm{\!a}, \boldsymbol{\psi}_\mathrm{\!b}}\!\right\}^{\!\!\intercal}\!$ is also positive semi-definite. Thus, we can establish
	\begin{equation}
		\begin{aligned}
			&\mathbb{E}_{\boldsymbol{\psi}_\mathrm{b}}\!\!\!\left\{\!\boldsymbol{\mathrm{J}}_{\boldsymbol{\psi}_\mathrm{a}, \boldsymbol{\psi}_\mathrm{a}}\!\right\} \geq \\ &\quad\quad\mathbb{E}_{\boldsymbol{\psi}_\mathrm{b}}\!\!\!\left\{\!\boldsymbol{\mathrm{J}}_{\boldsymbol{\psi}_\mathrm{a}, \boldsymbol{\psi}_\mathrm{a}}\!\right\} 
			\!-\! \mathbb{E}_{\boldsymbol{\psi}_\mathrm{b}}\!\!\!\left\{\!\boldsymbol{\mathrm{J}}_{\boldsymbol{\psi}_\mathrm{a}, \boldsymbol{\psi}_\mathrm{b}}\!\right\}\! \mathbb{E}_{\boldsymbol{\psi}_\mathrm{b}}\!\!\!\left\{\!\boldsymbol{\mathrm{J}}_{\boldsymbol{\psi}_\mathrm{b}, \boldsymbol{\psi}_\mathrm{b}}\!\right\}^{\!\!-\!1}\!\mathbb{E}_{\boldsymbol{\psi}_\mathrm{b}}\!\!\!\left\{\!\boldsymbol{\mathrm{J}}_{\boldsymbol{\psi}_\mathrm{a}, \boldsymbol{\psi}_\mathrm{b}}\!\right\}^{\!\!\intercal}\!. 
		\end{aligned}
	\end{equation}
	According to the monotone property of the matrix inverse \cite{Appendix_Matrix_analysis},
	\begin{equation}
		\label{equ_inverse_inequality}
		\begin{aligned}
			&\mathbb{E}_{\boldsymbol{\psi}_\mathrm{b}}\!\!\!\left\{\!\boldsymbol{\mathrm{J}}_{\boldsymbol{\psi}_\mathrm{a}, \boldsymbol{\psi}_\mathrm{a}}\!\right\}^{-1} \leq \\ &\,\,\left(\mathbb{E}_{\boldsymbol{\psi}_\mathrm{b}}\!\!\!\left\{\!\boldsymbol{\mathrm{J}}_{\boldsymbol{\psi}_\mathrm{a}, \boldsymbol{\psi}_\mathrm{a}}\!\right\} 
			\!-\! \mathbb{E}_{\boldsymbol{\psi}_\mathrm{b}}\!\!\!\left\{\!\boldsymbol{\mathrm{J}}_{\boldsymbol{\psi}_\mathrm{a}, \boldsymbol{\psi}_\mathrm{b}}\!\right\}\! \mathbb{E}_{\boldsymbol{\psi}_\mathrm{b}}\!\!\!\left\{\!\boldsymbol{\mathrm{J}}_{\boldsymbol{\psi}_\mathrm{b}, \boldsymbol{\psi}_\mathrm{b}}\!\right\}^{\!\!-\!1}\!\mathbb{E}_{\boldsymbol{\psi}_\mathrm{b}}\!\!\!\left\{\!\boldsymbol{\mathrm{J}}_{\boldsymbol{\psi}_\mathrm{a}, \boldsymbol{\psi}_\mathrm{b}}\!\right\}^{\!\!\intercal}\right)^{\!\!-\!1} \!\!. 
		\end{aligned}
	\end{equation}
	Applying the Schur complement to the expectation of the FIM over $\boldsymbol{\psi}_{\mathrm{b}}$, we can establish
	\begin{equation}
		\label{equ_shur_complement}
		\begin{aligned}
			&\left[\mathbb{E}_{\boldsymbol{\psi}_{\mathrm{b}}}\!\left\{\!\boldsymbol{\mathrm{J}}_{\boldsymbol{\psi}_{\mathrm{a}}}\right\}^{-1}\right]_{1:L_{\mathrm{a}},1:L_{\mathrm{a}}} = \\
			& \,\,\, \left(\mathbb{E}_{\boldsymbol{\psi}_\mathrm{b}}\!\!\!\left\{\!\boldsymbol{\mathrm{J}}_{\boldsymbol{\psi}_\mathrm{a}, \boldsymbol{\psi}_\mathrm{a}}\!\right\} 
			\!-\! \mathbb{E}_{\boldsymbol{\psi}_\mathrm{b}}\!\!\!\left\{\!\boldsymbol{\mathrm{J}}_{\boldsymbol{\psi}_\mathrm{a}, \boldsymbol{\psi}_\mathrm{b}}\!\right\}\! \mathbb{E}_{\boldsymbol{\psi}_\mathrm{b}}\!\!\!\left\{\!\boldsymbol{\mathrm{J}}_{\boldsymbol{\psi}_\mathrm{b}, \boldsymbol{\psi}_\mathrm{b}}\!\right\}^{\!\!-\!1}\!\mathbb{E}_{\boldsymbol{\psi}_\mathrm{b}}\!\!\!\left\{\!\boldsymbol{\mathrm{J}}_{\boldsymbol{\psi}_\mathrm{a}, \boldsymbol{\psi}_\mathrm{b}}\!\right\}^{\!\!\intercal}\right)^{\!\!-\!1} \!\!\! .
		\end{aligned}
	\end{equation}
	Substituting \eqref{equ_shur_complement} into \eqref{equ_inverse_inequality}, we can obtain 
	\begin{equation}
		\mathbb{E}_{\boldsymbol{\psi}_\mathrm{b}}\!\left\{\!\boldsymbol{\mathrm{J}}_{\boldsymbol{\psi}_\mathrm{a}, \boldsymbol{\psi}_\mathrm{a}}\right\}^{-1} \leq \left[\mathbb{E}_{\boldsymbol{\psi}_{\mathrm{b}}}\!\left\{\!\boldsymbol{\mathrm{J}}_{\boldsymbol{\psi}_{\mathrm{a}}}\right\}^{-1}\right]_{1:L_{\mathrm{a}},1:L_{\mathrm{a}}} ,
	\end{equation}
	which completes the proof.

\section{Derivation of the Dynamic Path AoA HRCRB}
\setcounter{equation}{0}
\setcounter{equation}{0}
	\label{appendix_CRB_theta}
		
	From the reordered FIM (19), 
	the entry $\mathrm{J}^{\text{S}}_{\theta_{\mathrm d}\!, \theta_{\mathrm d}}$ is given in \eqref{FIM_entries}, and submatrices $\boldsymbol{\mathrm{J}}^{\text{S}}_{\boldsymbol{\psi}^{\text{S}}_t\!,\boldsymbol{\psi}^{\text{S}}_t}\!$ and $\boldsymbol{\mathrm{J}}^{\text{S}}_{\theta_{\mathrm d}\!, \boldsymbol{\psi}^{\text{S}}_t}\!$ and are given by \eqref{equ_FIM_reorder_entity} at the bottom\stepcounter{equation}. 
	
	Dividing the re-ordered FIM in (19) into blocks, 
	\begin{equation}
		\begin{aligned}
		\boldsymbol{\mathrm{J}}^{\text{S}} & = 
			\begin{bmatrix}
				\begin{array}{c | c}
					\boldsymbol{\mathrm{A}} & \boldsymbol{\mathrm{B}}^{\intercal} \\  
					\hline
					\boldsymbol{\mathrm{B}} & \boldsymbol{\mathrm{C}}
				\end{array}
			\end{bmatrix} 
		 = \begin{bmatrix}
				\begin{array}{c | ccc}
				\mathrm{J}^{\text{S}}_{\theta_{\mathrm{d}},\theta_{\mathrm{d}}} & \boldsymbol{\mathrm{J}}^{\text{S}}_{\theta_{\mathrm{d}},\boldsymbol{\psi}_1^{\text{S}}} & \cdots & \boldsymbol{\mathrm{J}}^{\text{S}}_{\theta_{\mathrm{d}},\boldsymbol{\psi}_T^{\text{S}}}  \\  
				\hline
				{\boldsymbol{\mathrm{J}}^{\text{S}}_{\theta_{\mathrm{d}},\boldsymbol{\psi}_1^{\text{S}}}}^{\!\!\!\!\intercal} & \boldsymbol{\mathrm{J}}^{\text{S}}_{\boldsymbol{\psi}_1^{\text{S}},\boldsymbol{\psi}_1^{\text{S}}} & & \boldsymbol{0}\\
				\vdots & & \ddots & \\
				{\boldsymbol{\mathrm{J}}^{\text{S}}_{\theta_{\mathrm{d}},\boldsymbol{\psi}_T^{\text{S}}}}^{\!\!\!\!\intercal} & \boldsymbol{0} & & \boldsymbol{\mathrm{J}}^{\text{S}}_{{\boldsymbol{\psi}_T^{\text{S}},\boldsymbol{\psi}_T^{\text{S}}}}
			\end{array}
		\end{bmatrix},
		\end{aligned}
	\end{equation}
	the equivalent Fisher information of $\theta_{\mathrm{d}}$ is given by the Schur complement,
	\begin{equation}
		\mathrm{J}_{\theta_{\mathrm{d}}}^{\text{S},\mathrm{equ}} = \boldsymbol{\mathrm{A}} - 
		\boldsymbol{\mathrm{B}}^{\intercal} \boldsymbol{\mathrm{C}}^{-1} \boldsymbol{\mathrm{B}} . 
	\end{equation}
	Since $\boldsymbol{\mathrm{C}}$ is a block diagonal matrix,  
	\begin{equation}
		\label{equ_FIM_theta_equ}
		\begin{aligned}
			&\mathrm{J}_{\theta_{\mathrm{d}}}^{\text{S},\mathrm{equ}} \\
			&\quad= \mathrm{J}^{\text{S}}_{\theta_{\mathrm{d}},\theta_{\mathrm{d}}} \!-\! 
			\begin{bmatrix}
				\boldsymbol{\mathrm{J}}^{\text{S}}_{\theta_{\mathrm{d}},\boldsymbol{\psi}_1^{\text{S}}} \\
				\vdots \\
				\boldsymbol{\mathrm{J}}^{\text{S}}_{\theta_{\mathrm{d}},\boldsymbol{\psi}_T^{\text{S}}}
			\end{bmatrix}^{\!\!\intercal}\!\!
			\begin{bmatrix}
				{\boldsymbol{\mathrm{J}}^{\text{S}}_{\boldsymbol{\psi}_1^{\text{S}},\boldsymbol{\psi}_1^{\text{S}}}}^{\!\!\!\!\!\!-1} & &0 \\
				&\ddots & \\
				0 & &{\boldsymbol{\mathrm{J}}^{\text{S}}_{\boldsymbol{\psi}_T^{\text{S}},\boldsymbol{\psi}_T^{\text{S}}}}^{\!\!\!\!\!\!-1}
			\end{bmatrix}\!\!
			\begin{bmatrix}
				\boldsymbol{\mathrm{J}}^{\text{S}}_{\theta_{\mathrm{d}},\boldsymbol{\psi}_1^{\text{S}}} \\
				\vdots \\
				\boldsymbol{\mathrm{J}}^{\text{S}}_{\theta_{\mathrm{d}},\boldsymbol{\psi}_T^{\text{S}}}
			\end{bmatrix} \\
		&\quad= \mathrm{J}^{\text{S}}_{\theta_{\mathrm{d}},\theta_{\mathrm{d}}} \!-\! \sum_{t=1}^T \boldsymbol{\mathrm{J}}^{\text{S}}_{\theta_{\mathrm{d}},\boldsymbol{\psi}_t^{\text{S}}} {\boldsymbol{\mathrm{J}}^{\text{S}}_{\boldsymbol{\psi}_t^{\text{S}},\boldsymbol{\psi}_t^{\text{S}}}}^{\!\!\!\!\!\!-1} {\boldsymbol{\mathrm{J}}^{\text{S}}_{\theta_{\mathrm{d}},\boldsymbol{\psi}_t^{\text{S}}}}^{\!\!\!\!\intercal} .
		\end{aligned}
	\end{equation}
	The matrix inverse ${\boldsymbol{\mathrm{J}}^{\text{S}}_{\boldsymbol{\psi}_t^{\text{S}},\boldsymbol{\psi}_t^{\text{S}}}}^{\!\!\!\!\!\!-1}$ is given by 
	\eqref{equ_FIM_inverse} at the bottom of the page,
	where $\chi = \boldsymbol{a}_{\theta}\!\left(\theta_{\mathrm{d}}\right)^\dagger \boldsymbol{h}^{\text{S}}_{\mathrm{s}}+Md_t$.
	Substituting \eqref{equ_FIM_reorder_entity} and \eqref{equ_FIM_inverse} into \eqref{equ_FIM_theta_equ}, the expression of $\mathrm{J}_{\theta_{\mathrm{d}}}^{\text{S},\mathrm{equ}}$ is given by
	\eqref{equ_FIM_theta_equ_result} at the bottom of the page.
	Therefrom, by finding the expectation of \eqref{equ_FIM_theta_equ_result} over $\boldsymbol{d}$ and then finding the inverse, (24) is obtained, which completes the derivation of the HRCRB. 
	
	Inequality (30) holds according to Proposition 2.

\section{Proof of Theorem 1}
\setcounter{equation}{0}
	\label{appendix_proof_CRB_theta_property}
	For brevity, we denote $\boldsymbol{a}_{\theta}\!\left(\theta\right)$ and $\boldsymbol{b}_{\theta}\!\left(\theta\right)$ as $\boldsymbol{a}$ and $\boldsymbol{b}$ respectively.
	
	$\rho^{\text{S}}_{\text{\footnotesize $_\theta$}}$ can be expressed as
	\begin{equation}
		\rho^{\text{S}}_{\text{\footnotesize $_\theta$}}=\left(1-\Xi/\left(2\Gamma\Lambda\right)\right)^{-1},
	\end{equation}
	where $\Gamma \triangleq \left|\boldsymbol{a}\right|^2 \left|\boldsymbol{b}\right|^2 - |\boldsymbol{a}^\dagger \boldsymbol{b}|^2$, $\Delta \triangleq \left|\boldsymbol{a}\right|^2 \big|\boldsymbol{h}^{\text{S}}_{\mathrm s}\big|^2-|\boldsymbol{a}^\dagger \boldsymbol{h}^{\text{S}}_{\mathrm s}|^2$, and $\Xi \triangleq \big|\big(\boldsymbol{b}^\dagger\boldsymbol{a}\boldsymbol{a}^\dagger-\boldsymbol{a}^\dagger\boldsymbol{a}\boldsymbol{b}^\dagger\big)\boldsymbol{h}^{\text{S}}_\mathrm{s}\big|^2$.

	Apparently, $\Delta\geq 0, \Xi \geq 0$, and due to the geometry of the antenna array, $\Gamma > 0$. It is evident that $1 \leq \rho^{\text{S}}_{\text{\footnotesize $_\theta$}} $.
	
	The remaining part of the inequality in (27) is equivalent to
	\begin{equation}
		\label{equivalent}
		\rho^{\text{S}}_{\text{\footnotesize $_\theta$}} \leq2 \iff \Xi \leq \Gamma \Delta.
	\end{equation}
	
	According to the Binet–Cauchy identity, 
	\begin{equation}
		\Gamma=\!\!\!\!\sum_{1\leq i <j\leq M}\!\!\!\!\left(\boldsymbol{a}_i\boldsymbol{b}_j-\boldsymbol{a}_j\boldsymbol{b}_i\right)\left(\boldsymbol{a}_i^*\boldsymbol{b}_j^*-\boldsymbol{a}_j^*\boldsymbol{b}_i^*\right)=\left|\Lambda_{\boldsymbol{a}\boldsymbol{b}}\right|^2 /2 ,
	\end{equation}
	where $\Lambda_{\boldsymbol{a}\boldsymbol{b}}=\mathrm{vec}\left(\boldsymbol{a}\cdot \boldsymbol{b}^\intercal - \boldsymbol{b}\cdot \boldsymbol{a}^\intercal \right)$ . 
	
	Similarly, we have \mbox{$\Delta\!=\!\frac{1}{2}\left|\Lambda_{\boldsymbol{a}\boldsymbol{h}^{\text{S}}_{\mathrm{s}}}\right|^2$}  and \mbox{$\Xi\!=\!\frac{1}{4}\left|\Lambda_{\boldsymbol{a}\boldsymbol{h}^{\text{S}}_{\mathrm{s}}}^\intercal \! \cdot \Lambda_{\boldsymbol{b}\boldsymbol{a}}\right|^2 $}.
	
	Substituting $\Gamma$, $\Delta$, and $\Xi$ into (\ref{equivalent}), we equivalently require the following inequality is true:
	\begin{equation}
		\label{Equ-final-inequality}
		\big|\Lambda_{\boldsymbol{a}\boldsymbol{b}}^\dagger \cdot \Lambda_{\boldsymbol{a}\boldsymbol{h}^{\text{S}}_\mathrm{s}}\big|^2
		\leq \left|\Lambda_{\boldsymbol{a}\boldsymbol{b}}\right|^2 \left|\Lambda_{\boldsymbol{a}\boldsymbol{h}^{\text{S}}_\mathrm{s}}\right|^2.
	\end{equation}
	Equation \eqref{Equ-final-inequality} can be established via the Cauchy–Schwarz inequality, which completes the proof.

\section{Derivation of the Dynamic Path CGS AHRCRB}
\setcounter{equation}{0}
	\label{appendix_CRB_d}
	We first prove (29). 
	By moving the parameters corresponding to the $t$-th snapshot, $\boldsymbol{\psi}_t^{\text{S}}$, to the rear of the parameter vector, the corresponding FIM after reordering is given by
	\eqref{equ_FIM_CGS_blocks} at the bottom of the page.
	\begin{figure*}[b]
		\hrulefill
		\begin{equation}
			\label{equ_FIM_CGS_blocks}
			\begin{aligned}
				\boldsymbol{\mathrm{J}}^{\text{S}} & = \begin{bmatrix}
					\begin{array}{c | c}
						\boldsymbol{\mathrm{D}} & \boldsymbol{\mathrm{E}}^{\intercal} \\  
						\hline
						\boldsymbol{\mathrm{E}} & \boldsymbol{\mathrm{J}}_{{\boldsymbol{\psi}_t^{\text{S}},\boldsymbol{\psi}_t^{\text{S}}}}^{\text{S}}
					\end{array}
				\end{bmatrix} 
				 = \begin{bmatrix}
					\begin{array}{ccccccc | c}
						\mathrm{J}^{\text{S}}_{\theta_{\mathrm{d}},\theta_{\mathrm{d}}} &\boldsymbol{\mathrm{J}}^{\text{S}}_{\theta_{\mathrm{d}},\boldsymbol{\psi}_1^{\text{S}}} &\cdots &\boldsymbol{\mathrm{J}}^{\text{S}}_{\theta_{\mathrm{d}},\boldsymbol{\psi}_{t-1}^{\text{S}}} &\boldsymbol{\mathrm{J}}^{\text{S}}_{\theta_{\mathrm{d}},\boldsymbol{\psi}_{t+1}^{\text{S}}} &\cdots &\boldsymbol{\mathrm{J}}^{\text{S}}_{\theta_{\mathrm{d}},\boldsymbol{\psi}_T^{\text{S}}} &\boldsymbol{\mathrm{J}}^{\text{S}}_{\theta_{\mathrm{d}},\boldsymbol{\psi}_t^{\text{S}}} \\  
						{\boldsymbol{\mathrm{J}}^{\text{S}}_{\theta_{\mathrm{d}},\boldsymbol{\psi}_1^{\text{S}}}}^\intercal &\boldsymbol{\mathrm{J}}^{\text{S}}_{\boldsymbol{\psi}_1^{\text{S}},\boldsymbol{\psi}_1^{\text{S}}} & & & & & &\boldsymbol{0} \\  
						\vdots & &\ddots & & &\boldsymbol{0} & & \\  
						{\boldsymbol{\mathrm{J}}^{\text{S}}_{\theta_{\mathrm{d}},\boldsymbol{\psi}_{t-1}^{\text{S}}}}^\intercal  & & & \!\!\!\! \boldsymbol{\mathrm{J}}^{\text{S}}_{\boldsymbol{\psi}_{t-1}^{\text{S}},\boldsymbol{\psi}_{t-1}^{\text{S}}} \!\!\!\! & & & &\vdots \\  
						{\boldsymbol{\mathrm{J}}^{\text{S}}_{\theta_{\mathrm{d}},\boldsymbol{\psi}_{t+1}^{\text{S}}}}^\intercal  & & & & \!\!\!\! \boldsymbol{\mathrm{J}}^{\text{S}}_{\boldsymbol{\psi}_{t+1}^{\text{S}},\boldsymbol{\psi}_{t+1}^{\text{S}}} \!\!\!\! & & & \\  
						\vdots & &\boldsymbol{0} & & &\ddots & & \\ 
						{\boldsymbol{\mathrm{J}}^{\text{S}}_{\theta_{\mathrm{d}},\boldsymbol{\psi}_T^{\text{S}}}}^\intercal & & & & & &\boldsymbol{\mathrm{J}}^{\text{S}}_{\boldsymbol{\psi}_T^{\text{S}},\boldsymbol{\psi}_T^{\text{S}}} &\boldsymbol{0} \\ 
						\hline
						{\boldsymbol{\mathrm{J}}^{\text{S}}_{\theta_{\mathrm{d}},\boldsymbol{\psi}_t^{\text{S}}}}^\intercal & \boldsymbol{0} & &\cdots & & &\boldsymbol{0} &\boldsymbol{\mathrm{J}}^{\text{S}}_{{\boldsymbol{\psi}_t^{\text{S}},\boldsymbol{\psi}_t^{\text{S}}}}
					\end{array}
				\end{bmatrix},
			\end{aligned}
		\end{equation}
	\end{figure*}
	Dividing the FIM into blocks as shown in \eqref{equ_FIM_CGS_blocks}, the EFIM of $\boldsymbol{\psi}_t^{\text{S}}$ is given by
	\begin{equation}
		\label{Equ_J_psi_kappa_equ}
		\begin{aligned}
		\boldsymbol{\mathrm{J}}_{\boldsymbol{\psi}_t^{\text{S}}}^{\text{S},\mathrm{equ}} &=\boldsymbol{\mathrm{J}}^{\text{S}}_{\boldsymbol{\psi}_t^{\text{S}},\boldsymbol{\psi}_t^{\text{S}}}-\boldsymbol{\mathrm{E}} \boldsymbol{\mathrm{D}}^{-1} \boldsymbol{\mathrm{E}}^{\intercal} \\
		&=\boldsymbol{\mathrm{J}}^{\text{S}}_{\boldsymbol{\psi}_t^{\text{S}},\boldsymbol{\psi}_t^{\text{S}}}-{\boldsymbol{\mathrm{J}}^{\text{S}}_{\theta,\boldsymbol{\psi}_t^{\text{S}}}}^{\!\!\!\!\intercal} \left[\boldsymbol{\mathrm{D}}^{-1}\right]_{1,1} \boldsymbol{\mathrm{J}}^{\text{S}}_{\theta,\boldsymbol{\psi}_t^{\text{S}}} . 
		\end{aligned}
	\end{equation}
	Further applying Schur complement to $\left[\boldsymbol{\mathrm{D}}^{-1}\right]_{1,1}$, and following a line of reasoning similar to that in \eqref{equ_FIM_theta_equ}, we obtain
	\begin{equation}
		\left[\boldsymbol{\mathrm{D}}^{-1}\right]_{1,1} = \Big( 
		\mathrm{J}^{\text{S}}_{\theta,\theta}- \!\!\!\!\!\! \sum_{\iota \in [1 \ldots T] \backslash \{t\}} \!\!\!\!\!\! \boldsymbol{\mathrm{J}}^{\text{S}}_{\theta,\boldsymbol{\psi}_\iota^{\text{S}}} {\boldsymbol{\mathrm{J}}^{\text{S}}_{\boldsymbol{\psi}_\iota^{\text{S}},\boldsymbol{\psi}_\iota^{\text{S}}}}^{\!\!\!\!-1} {\boldsymbol{\mathrm{J}}^{\text{S}}_{\theta,\boldsymbol{\psi}_\iota^{\text{S}}}}^{\!\!\!\!\intercal} 
		\Big)^{-1} .
	\end{equation}
	Thus, $\boldsymbol{\mathrm{J}}_{\boldsymbol{\psi}_t^{\text{S}}}^{\text{S},\mathrm{equ}}$ is given by
	\begin{equation}
		\label{Equ_FIM_psi_k}
		\begin{aligned}
			& \boldsymbol{\mathrm{J}}_{\boldsymbol{\psi}_t^{\text{S}}}^{\text{S},\mathrm{equ}} = 
			\boldsymbol{\mathrm{J}}^{\text{S}}_{\boldsymbol{\psi}_t^{\text{S}},\boldsymbol{\psi}_t^{\text{S}}} \\
			&- {\boldsymbol{\mathrm{J}}^{\text{S}}_{\theta,\boldsymbol{\psi}_t^{\text{S}}}}^{\!\!\!\!\intercal}  
			\Big( 
			\mathrm{J}^{\text{S}}_{\theta,\theta}- \!\!\!\!\!\! \sum_{\iota \in [1 \ldots T] \backslash \{t\}} \!\!\!\!\!\! \boldsymbol{\mathrm{J}}^{\text{S}}_{\theta,\boldsymbol{\psi}_\iota^{\text{S}}} {\boldsymbol{\mathrm{J}}^{\text{S}}_{\boldsymbol{\psi}_\iota^{\text{S}},\boldsymbol{\psi}_\iota^{\text{S}}}}^{\!\!\!\!-1} {\boldsymbol{\mathrm{J}}^{\text{S}}_{\theta,\boldsymbol{\psi}_\iota^{\text{S}}}}^{\!\!\!\!\intercal}
			\Big)^{-1} 
			\boldsymbol{\mathrm{J}}^{\text{S}}_{\theta,\boldsymbol{\psi}_t^{\text{S}}}.
		\end{aligned}
	\end{equation}

	Define vector $\boldsymbol{d}_{\backslash t}=[d_1,\ldots,d_{t-1},d_{t+1}\ldots,d_T]^{\intercal}$. The expectation of $\left[\boldsymbol{\mathrm{D}}^{-1}\right]_{1,1}$ over $\boldsymbol{d}_{\backslash t}$ is closely related to the dynamic path AoA HRCRB, 
	\begin{subequations}
		\begin{align}
			&\mathbb{E}_{\boldsymbol{d}_{\backslash t}}\left\{ \left[\boldsymbol{\mathrm{D}}^{-1}\right]_{1,1} \right\}\\
			&= \mathbb{E}_{\boldsymbol{d}_{\backslash t}}\Bigg\{ \!\! \Big( 
			\mathrm{J}^{\text{S}}_{\theta_{\mathrm{d}},\theta_{\mathrm{d}}}- \!\!\!\!\!\! \sum_{\iota \in [1 \ldots T] \backslash \{t\}} \!\!\!\!\!\! \boldsymbol{\mathrm{J}}^{\text{S}}_{\theta_{\mathrm{d}},\boldsymbol{\psi}_\iota^{\text{S}}} {\boldsymbol{\mathrm{J}}^{\text{S}}_{\boldsymbol{\psi}_\iota^{\text{S}},\boldsymbol{\psi}_\iota^{\text{S}}}}^{\!\!\!\!-1} {\boldsymbol{\mathrm{J}}^{\text{S}}_{\theta_{\mathrm{d}},\boldsymbol{\psi}_\iota^{\text{S}}}}^{\!\!\!\!\intercal}
			\Big)^{\!\!-1} \! \Bigg\} \label{Equ_D_inequ:a}\\
			&= \mathbb{E}_{\boldsymbol{d}_{\backslash t}}\Bigg\{ \!\! \bigg( {\mathrm{J}_{\theta_{\mathrm{d}}}^{\text{S},\mathrm{equ}, {\backslash t}}} \notag \\
			& \quad \!+\! \frac{{\left| d_{t}\right|}^{2}}{{\sigma}^{2}M} \!\left( {|\boldsymbol{a}_{\theta}\!(\theta_{\mathrm{d}})|}^{2}{|\boldsymbol{b}(\theta_{\mathrm{d}})|}^{2}- {|{\boldsymbol{a}_{\theta}\!(\theta_{\mathrm{d}})}^{\dagger} \boldsymbol{b}(\theta_{\mathrm{d}})|}^{2} \right) \!\!\!
			\bigg)^{\!\! -1} \!\Bigg\} \label{Equ_D_inequ:b}\\
			&\leq \mathbb{E}_{\boldsymbol{d}_{\backslash t}} \left\{{\mathrm{J}_{\theta_{\mathrm{d}}}^{\text{S},\mathrm{equ}, {\backslash t}}}^{-1}\right\}, \label{Equ_D_inequ:c}
		\end{align}
	\end{subequations}
	where $\mathbb{E}_{\boldsymbol{d}_{\backslash t}}\!\! \left\{{\mathrm{J}_{\theta_{\mathrm{d}}}^{\text{S},\mathrm{equ}, {\backslash t}}}^{-1}\right\}$ in \eqref{Equ_D_inequ:c} is in accordance with the CRB of dynamic path AoA estimation, $\mathbb{E}_{\boldsymbol{d}} \left\{{\mathrm{J}_{\theta_{\mathrm{d}}}^{\text{S},\mathrm{equ}}}^{-1}\right\}$, except that the $t$-th snapshot is excluded from the observation. 
	As $T$ tends to infinity, $\mathbb{E}_{\boldsymbol{d}_{\backslash t}}\!\! \left\{{\mathrm{J}_{\theta_{\mathrm{d}}}^{\text{S},\mathrm{equ}, {\backslash t}}}^{-1}\!\right\}$ tends to zero, and \mbox{$\lim\limits_{\!\!T \to +\infty\!\!} \!\!\mathbb{E}_{\boldsymbol{d}_{\backslash t}}\!\! \left\{{\mathrm{J}_{\theta_{\mathrm{d}}}^{\text{S},\mathrm{equ}, {\backslash t}}}^{-1}\!\right\}\!=\!0$}. Finding the limit of \eqref{Equ_D_inequ:c} as $T$ tends to infinity, 
	\begin{equation}
		0 
		\leq \lim_{T \to +\infty} \!\! \mathbb{E}_{\boldsymbol{d}_{\backslash t}} \!\! \left\{ \left[\boldsymbol{\mathrm{D}}^{-1}\right]_{1,1} \right\} 
		\leq \lim_{T \to +\infty} \!\! \mathbb{E}_{\boldsymbol{d}_{\backslash t}} \!\! \left\{{\mathrm{J}_{\theta_{\mathrm{d}}}^{\text{S},\mathrm{equ}, \boldsymbol{\backslash t}}}^{-1}\right\},
	\end{equation}
	thus, $\lim\limits_{T \to + \infty} \!\!\mathbb{E}_{\boldsymbol{d}_{\backslash t}}\!\!\left\{ \left[\boldsymbol{\mathrm{D}}^{-1}\right]_{1,1} \right\}=0$.
	
	Finding the expectation of \eqref{Equ_J_psi_kappa_equ} over $\boldsymbol{d}$ and then the limit as $T$ tends to infinity, the asymptotic EFIM of $\boldsymbol{\psi}_{t}^{\text{S}}$ is given by 
	\begin{equation}
		\begin{aligned}
			&\boldsymbol{\mathrm{J}}_{\text{S},\boldsymbol{\psi}_{t}^{\text{S}}}^{\mathrm{equ, asym}} \\
			&= \lim_{T \to +\infty} \mathbb{E}_{\boldsymbol{d}}\left\{ \boldsymbol{\mathrm{J}}_{\boldsymbol{\psi}_{t}^{\text{S}}}^{\text{S},\mathrm{equ}} \right\}\\
			&= \mathbb{E}_{{d}_{t}}\!\!\left\{ \lim_{T \to +\infty} \mathbb{E}_{\boldsymbol{d}_{\backslash t}}\left\{ \boldsymbol{\mathrm{J}}_{\boldsymbol{\psi}_{t}^{\text{S}}}^{\text{S},\mathrm{equ}} \right\} \right\}\\
			&= \mathbb{E}_{{d}_{t}}\!\!\left\{ \boldsymbol{\mathrm{J}}^{\text{S}}_{\boldsymbol{\psi}_{t}^{\text{S}}, \boldsymbol{\psi}_{t}^{\text{S}}} \!-\! {\boldsymbol{\mathrm{J}}^{\text{S}}_{\theta_{\mathrm{d}}, \boldsymbol{\psi}_{t}^{\text{S}}}}^{\!\!\!\!\intercal} \boldsymbol{\mathrm{J}}^{\text{S}}_{\theta_{\mathrm{d}}, \boldsymbol{\psi}_{t}^{\text{S}}} \!\! \cdot \!\! \lim_{T \to + \infty} \!\! \mathbb{E}_{\boldsymbol{d}_{\backslash t}}\!\!\left\{ \left[\boldsymbol{\mathrm{D}}^{-1}\right]_{1,1} \right\} \!\! \right\}\\
			&= \mathbb{E}_{{d}_{t}}\!\!\left\{ \boldsymbol{\mathrm{J}}^{\text{S}}_{\boldsymbol{\psi}_{t}^{\text{S}}, \boldsymbol{\psi}_{t}^{\text{S}}}\right\}, 
		\end{aligned}
	\end{equation}
	which completes the proof of (29).
	
	Then, we prove (31). 
	Substitute $\!\!\lim\limits_{T \to +\infty} \!\!\!\mathbb{E}_{\boldsymbol{d}}\!\left\{\boldsymbol{\mathrm{J}}_{\boldsymbol{\psi}^{\text{S}}_t}^{\text{S},\mathrm{equ}} \right\}$ from
	(29) into (30), 
	\begin{equation}
		\begin{aligned}
		\label{Equ_CRB_d_derive}
		&\text{AHRCRB}^{\text{S}}_{\mathrm{d}}\\ 
		&= \!\!\!
		\lim_{T \to +\infty} \frac{1}{T} \!\cdot\! \mathbb{E}_{\boldsymbol{d}}\!\left\{ 
		\sum_{t=1}^{T} \mathrm{Tr}\!\! \left(\!\left[
		{\boldsymbol{\mathrm{J}}_{\boldsymbol{\psi}^{\text{S}}_t}^{\text{S},\mathrm{equ}} }^{-1} \right]_{\!1:2,1:2} \!\right)\!\! \right\} \\
		&= \!\!\!\lim_{T \to +\infty} \frac{1}{T} \sum_{t=1}^{T} \mathrm{Tr}\! \left(  \left[
		{\left(\lim_{T \to +\infty} \!\!\!\mathbb{E}_{\boldsymbol{d}}\!\left\{\boldsymbol{\mathrm{J}}_{\boldsymbol{\psi}^{\text{S}}_t}^{\text{S},\mathrm{equ}} \right\}\right)}^{\!\!-1} \right]_{\!1:2,1:2}  \right)\\
		&= \!\!\!\lim_{T \to +\infty} \frac{1}{T} \mathbb{E}_{\boldsymbol{d}}\!\!\left\{ \sum_{t=1}^{T} \mathrm{Tr}\! \left(  \left[
		{\boldsymbol{\mathrm{J}}^{\text{S}}_{\boldsymbol{\psi}_{t}^{\text{S}}, \boldsymbol{\psi}_{t}^{\text{S}}}}^{-1} \right]_{\!1:2,1:2}  \right)\right\}
		\vspace{-12pt}
		\end{aligned}
	\end{equation}
	where ${\boldsymbol{\mathrm{J}}^{\text{S}}_{\boldsymbol{\psi}_t^{\text{S}},\boldsymbol{\psi}_t^{\text{S}}}}^{\!\!\!\!-1}$ is given in \eqref{equ_FIM_inverse}. Further substitute ${\boldsymbol{\mathrm{J}}^{\text{S}}_{\boldsymbol{\psi}_t^{\text{S}},\boldsymbol{\psi}_t^{\text{S}}}}^{\!\!\!\!-1}$ from \eqref{equ_FIM_inverse} into \eqref{Equ_CRB_d_derive}, and due to the circularly symmetric distribution of the entries of $\boldsymbol{d}$,
	\begin{equation}
		\begin{aligned}
			&\text{AHRCRB}^{\text{S}}_{\mathrm{d}} \\
			&= \!\!\!\lim_{T \to +\infty} \frac{1}{T}\! \sum_{t=1}^{T} \mathbb{E}_{d_{t}}\!\!\!\left\{\!\!
			\frac{2\sigma^2}{M}\!+\!\frac{\sigma^2}{M} \!\cdot\! \frac{\Big|\boldsymbol{a}_{\theta}\!\left(\theta_{\mathrm{d}}\right)^\dagger \boldsymbol{h}^{\text{S}}_\mathrm{s}+M \boldsymbol{d}_{t}\Big|^2}{\Big|\boldsymbol{a}_{\theta}\!\left(\theta_{\mathrm{d}}\right)\Big|^{\!2\!} \Big|\boldsymbol{h}^{\text{S}}_\mathrm{s}\Big|^2 \!\!-\!\Big|\boldsymbol{a}_{\theta}\!\left(\theta_{\mathrm{d}}\right)^\dagger\! \boldsymbol{h}^{\text{S}}_\mathrm{s}\Big|^2}\!\!\right\}\\
			&= \frac{2\sigma^2}{M}+\frac{\sigma^2}{M} \!\cdot\! \frac{\Big|\boldsymbol{a}_{\theta}\!\left(\theta_{\mathrm{d}}\right)^\dagger \boldsymbol{h}^{\text{S}}_\mathrm{s}\Big|^2+M^2 P_\mathrm{d}}{\Big|\boldsymbol{a}_{\theta}\!\left(\theta_{\mathrm{d}}\right)\Big|^2 \Big|\boldsymbol{h}^{\text{S}}_\mathrm{s}\Big|^2 -\Big|\boldsymbol{a}_{\theta}\!\left(\theta_{\mathrm{d}}\right)^\dagger \boldsymbol{h}^{\text{S}}_\mathrm{s}\Big|^2} ,
		\end{aligned}
	\end{equation}

	Next, we prove (32). 
	According to the classic Cram\'{e}r-Rao inequality, 
	\begin{equation}
		\mathbb{E}_{\boldsymbol{\epsilon}}\Big\{\big|\hat{\boldsymbol{d}}-\boldsymbol{d}\big|^2\Big\} \geq \sum_{t=1}^{T}\mathrm{Tr}\left(\left[{\boldsymbol{\mathrm J}_{\boldsymbol{\psi}_t^{\text{S} }}^{\text{S},\mathrm{equ}}}^{-1}\right]_{1:2,1:2}\right). 
	\end{equation}
	Find the expectation over $\boldsymbol{d}$, divide both sides by $T$, and find the limit as $T$ tends to infinity,
	\begin{equation}
		\begin{aligned}
			&\lim_{T\to+\infty}\frac{1}{T}\mathbb{E}_{\boldsymbol{\epsilon}, \boldsymbol{d}}\Big\{\big|\hat{\boldsymbol{d}}-\boldsymbol{d}\big|^{\!2}\Big\} \\
			&\quad\quad\geq  \lim_{T\to+\infty}\frac{1}{T}\mathbb{E}_{\boldsymbol{d}}\bigg\{\sum_{t=1}^{T}\mathrm{Tr}\left(\left[{\boldsymbol{\mathrm J}_{\boldsymbol{\psi}_t^{\text{S} }}^{\text{S},\mathrm{equ}}}^{-1}\right]_{1:2,1:2}\right)\bigg\}. 
		\end{aligned}
	\end{equation}
	Thus, 
	\begin{equation}
		\lim_{T \to \infty} \text{MSE}_{\mathrm d} \geq \text{AHRCRB}_{\mathrm d}^{\text{S}},
	\end{equation}
	which completes the proof.

\section{\blue{Compared Dynamic Path CGS Estimation Algorithm}}
\setcounter{equation}{0}
\label{Appendix_CGS_estimation_alogrithm}
The detailed procedures of the compared dynamic path CGS estimation algorithm are as follow:

\indent \emph{Step 1. AoA Estimation}: Obtain the AoA parameter estimation $\hat \theta$ using the MUSIC algorithm.
\\
\indent \emph{Step 2. Beam space Division}: Divide the beam space $\mathbb{C}^M$ into two orthogonal subspaces spanned by column vector sets $\boldsymbol{A}$ and $\boldsymbol{B}$ respectively.  $\boldsymbol{A}$ is constructed directly with the streering vector of the estimated AoA, $\boldsymbol{A} \leftarrow \nicefrac{\boldsymbol{a}_{\theta}\!(\hat \theta)}{\left|\boldsymbol{a}_{\theta}\!(\hat \theta)\right|}$; apply singular value decomposition to $\boldsymbol{A}$, $\boldsymbol{A} \! = \! \boldsymbol{\mathrm{U}}\boldsymbol{\Sigma}\boldsymbol{\mathrm{V}}^{\dagger}$, and $\boldsymbol{B} \! \in \! \mathbb{C}^{M \times (M-1)}$ is composed of the columns of $\boldsymbol{\mathrm{U}}$ that corresponds to the $M-1$ zero singular values. Consequently, $\boldsymbol{A}$ spans the subspace containing the dynamic path component, and $\boldsymbol{B}$ spans the nullspace of the dynamic path. 
\\
\indent \emph{Step 3. Dynamic Component Mitigation}: Mitigate the dynamic path component by projecting the array signal into the nullspace, $\boldsymbol{H}_{\mathrm{p}} \!\leftarrow\! \boldsymbol{B}^\dagger \boldsymbol{H}^{\text{S}}$. The result is the projection of the static path signals, $\boldsymbol{H}_{\mathrm{p}}\!=\!\!\boldsymbol{B}^\dagger \boldsymbol{h}^{\text{S}}_\mathrm{s}\cdot\mathrm{e}^{\mathrm{j}\boldsymbol{\varphi}_{\mathrm o}^\intercal} + \boldsymbol{n}$, where $\boldsymbol{n}$ is the i.i.d. Gaussian noise. 
\\
\indent \emph{Step 4. Random Phase Offset Estimation}: Apply maximal-ratio combining to the projection from the $M-1$ dimensions of the nullspace, $\boldsymbol{h}_\mathrm{q} \leftarrow \boldsymbol{w}^\dagger \boldsymbol{H}_{\mathrm{p}}$, where $\boldsymbol{w}$ is the complex weight vector. Apply eigenvalue decomposition to the correlation matrix $\frac{1}{K}{\boldsymbol{H}_\mathrm{p}}^\dagger \boldsymbol{H}_{\mathrm{p}}$, and $\boldsymbol{w}$ is given by the normalized eigenvector corresponding to the non-zero eigenvalue. The combined projection can be seen as an estimation of the differential random phase offset in the exponential form, $\boldsymbol{h}_\mathrm{q}= h_{\mathrm{p}} \cdot \mathrm{e}^{\mathrm{j}\boldsymbol{\varphi}_{\mathrm o}^\intercal}+\boldsymbol{n}$, where $h_{\mathrm{p}}$ is a complex scalar. The differential random phase offset is given by $\hat{\boldsymbol{\varphi}_{\mathrm o}}\leftarrow \angle{\boldsymbol{h}_\mathrm{q}}$. 
To implement the constraint (7b), $\hat{\boldsymbol{\varphi}_{\mathrm o}} \leftarrow \hat{\boldsymbol{\varphi}_{\mathrm o}}-\frac{1}{T}\boldsymbol{1}_{T\times T}\cdot \hat{\boldsymbol{\varphi}_{\mathrm o}}$.
\\
\indent \emph{Step 5. Phase Compensation}: Compensate the random phase offset, $\boldsymbol{H}_{\mathrm{c}} \leftarrow \boldsymbol{H}^{\text{S}} \cdot \mathrm{diag}\left(\mathrm{e}^{-\mathrm{j}\cdot\hat{\boldsymbol{\varphi}_{\mathrm o}}}\right)$.
\\
\indent \emph{Step 6. Dynamic Component Combining}: 
Combine the signal from the beam space, $\hat{\boldsymbol{d}} \leftarrow \boldsymbol{A}^\dagger \boldsymbol{H}_{\mathrm{c}}$, then remove the DC component to implement the constraint (7a), $\hat{\boldsymbol{d}} \leftarrow \hat{\boldsymbol{d}} - \frac{1}{T} \boldsymbol{1}_{T\times T} \hat{\boldsymbol{d}}$.

\end{appendices}

\bibliographystyle{ieeetr}
\bibliography{ciations}
\vfill

\end{document}